\documentclass[aps,prx,reprint,superscriptaddress]{revtex4-2}

\usepackage{graphicx}
\usepackage{epstopdf}
\usepackage{dcolumn}
\usepackage{bm}
\usepackage{color} 
\usepackage{ulem}
\usepackage{soul}
\usepackage{xspace}
\usepackage{multirow}
\usepackage[hypertexnames=false]{hyperref}
\usepackage{physics}
\usepackage{here}
\usepackage{titlesec}
\usepackage[caption=false]{subfig}
\usepackage{txfonts}
\usepackage{amsfonts}
\usepackage{color}%
\definecolor{myorange}{rgb}{0.7,0.5,0.0}
\definecolor{mygreen}{rgb}{0.0,0.7,0.0}
\definecolor{purple}{rgb}{0.75,0.0,1.0}

\newcommand{\QPEC}{Quantum-Phase Electronics Center, University of Tokyo, Tokyo 113-8656, Japan}
\newcommand{\RCAST}{Research Center for Advanced Science and Technology, University of Tokyo, Tokyo 153-8904, Japan}
\newcommand{\RIKEN}{RIKEN Center for Emergent Matter Science, 2-1 Hirosawa, Wako, 351-0198, Japan}

\titleformat*{\section}{\normalsize\bfseries\sffamily\large\raggedright}
\titleformat*{\subsection}{\normalsize\bfseries\sffamily\raggedright}
\renewcommand\thesection{\arabic{section}}
\renewcommand\thesubsection{\arabic{subsection}}

\makeatletter
    \def\@seccntformat#1{\@ifundefined{#1@cntformat}%
       {\csname the#1\endcsname\space}
       {\csname #1@cntformat\endcsname}}
    \def\section@cntformat{\thesection.\space}
    \def\subsection@cntformat{\thesection.\thesubsection.\space} 
\makeatother

\renewcommand{\figurename}{\textsf{\textbf{Figure }}}
\renewcommand{\thefigure}{\textsf{\textbf{{\arabic{figure}}}}}

\definecolor{green}{rgb}{0,0.6,0.1}

\begin{document}

\title{
Topological Band Inversion and Chiral Majorana Mode in Hcp Thallium
}

\author{Motoaki Hirayama}%
\email{hirayama@ap.t.u-tokyo.ac.jp}
\affiliation{\QPEC}%
\affiliation{\RIKEN}%

\author{Takuya Nomoto}%
\email{nomoto@ap.t.u-tokyo.ac.jp}
\affiliation{\RCAST}

\author{Ryotaro Arita}%
\email{arita@ap.t.u-tokyo.ac.jp }
\affiliation{\RCAST}
\affiliation{\RIKEN}

\date{\today}

\begin{abstract}
\textsf{\textbf{
The chiral Majorana fermion is an exotic particle that is its own antiparticle.
It can arise in a one-dimensional edge of topological materials, and especially that in a topological superconductor can be exploited in non-Abelian quantum computation.
While the chiral Majorana mode (CMM) remains elusive, a promising situation is realized when superconductivity coexists with a topologically non-trivial surface state.
Here, we perform fully non-empirical calculation for the CMM considering superconductivity and surface relaxation, and show that hexagonal close-packed thallium (Tl) has an ideal electronic state that harbors the CMM.
The $k_z=0$ plane of Tl is a mirror plane, realizing a full-gap band inversion corresponding to a topological crystalline insulating phase.
Its surface and hinge are stable and easy to make various structures.
Another notable feature is that the surface Dirac point is very close to the Fermi level, so that a small Zeeman field can induce a topological transition.
Our calculation indicates that Tl will provide a new platform of the Majorana fermion.
}}
\end{abstract}

\keywords{Chiral Majorana fermion, topological material, superconductivity, elemental thallium, first-principles calculations.}


\maketitle


\section{Introduction}
The chiral Majorana mode (CMM) is a one-dimensional edge state at zero energy that may emerge in topological phases~\cite{Wilczek09,Alicea12,Beenakker13,Elliott15,Chiu18}.
The Majorana mode is its own antiparticle and enables quantum computing using non-Abelian statistics~\cite{Kitaev97,Kitaev03,Kitaev06,Nayak08,Aasen16,10.1063/5.0102999,PRXQuantum.4.020329}.
Materials realization of the CMM is of great interest, and a variety of possibilities have been investigated.
While a bulk spinless $p+ip$ superconductor can host the CMM, $p$-type superconductors are usually fragile against impurities
and there are not so many candidate materials~\cite{Green00,Volovik99,Ivanov01,Senthil00,Stone04,Sato16,Sato17}.
Another intriguing possibility is artificial heterostructures in which a situation similar to that of $p+ip$ superconductors is realized~\cite{Sato03,Fu08,Fu09,Akhmerov09,Braunecker10,Potter10,Lutchyn10,Qi10,Yuval10,Williams12,Wang2012,Mourik12,Potter12,Takei13,Nadj-Perge13,Nadj-Perge14,Xu14,Xu15,Rontynen15,Li16,Sun16,Gerbold17,He17,Jeon17,Rachel17,Lian18,Chen18,Lutchyn2018,Desjardins19,Palacio-Morales19,Gerbold19,Jack19,Kayyalha2020,Manna20,Prada20,Shen20,Xie21}.
In recent years, 
it has been found that some superconductors
such as $\beta $-PdBi$_2$~\cite{Sakano15}, Fe(Te$_{0.55}$Se$_{0.45}$)~\cite{Yin2015,Wang15,Zhang18,Wang18,Machida19,Zhang2019,Kong2019a,Kong2019b,Zhu2020,Liu2020,Chen2019,Chiu2020,Wang2020}, PbTaSe$_2$~\cite{Ali14,Bian16}, PdTe$_2$~\cite{Noh17}, $\beta$-RhPb$_2$~\cite{ZhangJF18}, A15 superconductors~\cite{Kim18}, and Heusler alloys~\cite{Guo18},
have Dirac dispersion on the surface. 
If the proximity effect opens a superconducting gap in the Dirac dispersion in such a system, the Majorana fermion could emerge within a single material.
The emergence of the Majorana corner state in higher-order topological superconductors has also been proposed~\cite{PhysRevLett.121.186801,PhysRevB.98.245413,PhysRevB.100.054513}.

Unlike the Majorana bound state (MBS) in the vortex~\cite{PhysRevB.44.9667,Volovik99,Hosur11,Chiu12,PhysRevB.107.214518,Li2022,Valentini2022,DasSarma2023,PhysRevLett.130.106001}, the CMM is a one-dimensional state determined by the surface topology in a superconducting phase.
For the CMM, 
the crystal structure is desirable to be as simple as possible.
Also, the surface must be stable.
Furthermore, the Dirac point at the surface must be located near the Fermi level.
The transition temperature should not be too low.
Although the proposals of the materials so far are very interesting and nontrivial,
unfortunately the systems do not meet some or all of these realistic conditions for the Majorana fermions.
While Fe(Te$_{0.55}$Se$_{0.45}$) is promising in terms of the above requirements,
it comprises three elements and adjustment of the composition ratio is necessary.
Experimental results suggest that the coexistence of vortices with and without the MBS is 
related to quenched disorder~\cite{Machida19},
which may have more significant impact on the CMM.
Ultimately it is desirable to realize the CMM in a single element because of the simplicity.

\begin{figure}[ptb]
	\centering 
	\includegraphics[clip,width=0.45\textwidth ]{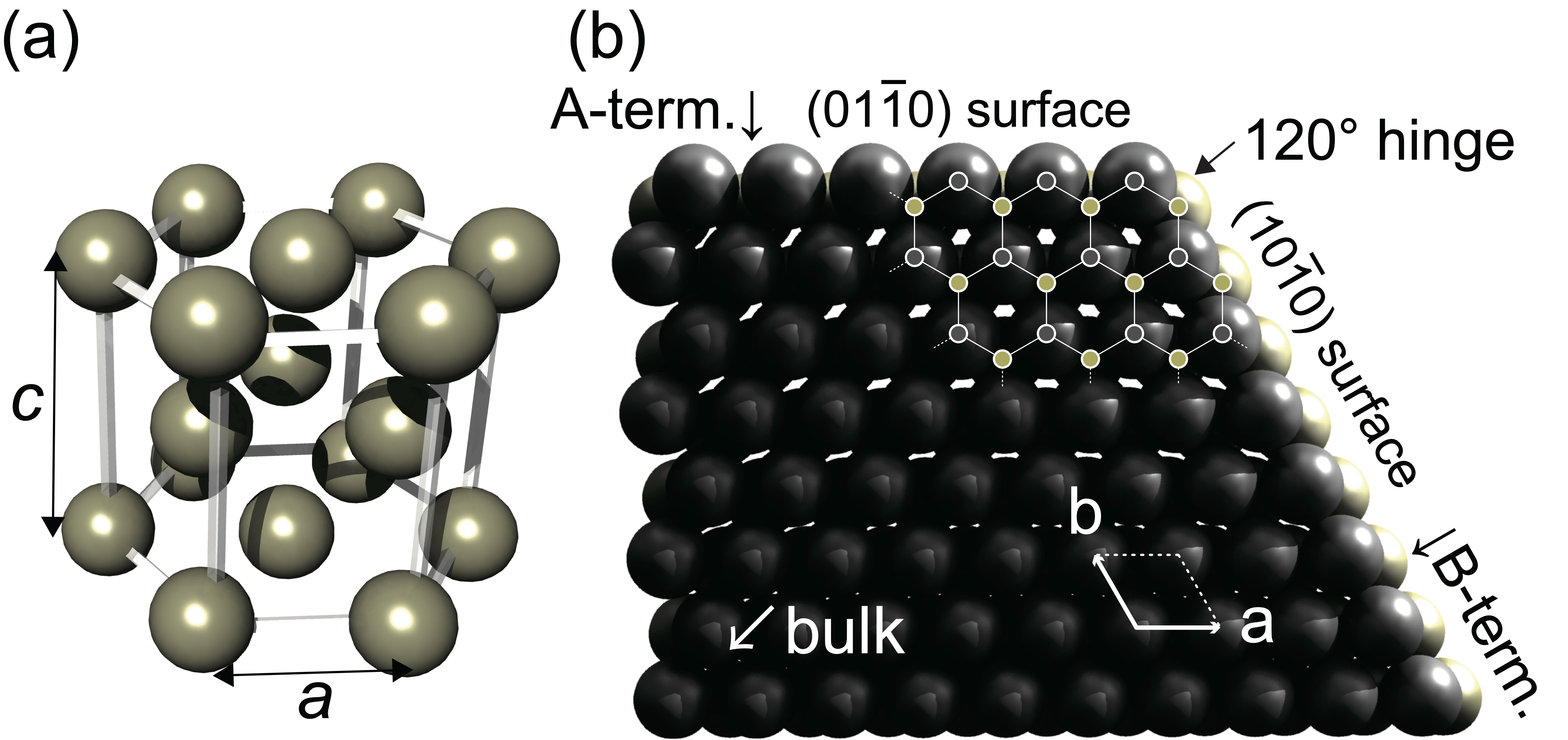} 
	\caption{Bulk and surface structure.
		(a) Crystal structure of hcp Tl.
		(b) Hinge of Tl between A-terminated $(01\bar{1}0)$ and B-terminated $(10\bar{1}0)$ surfaces.
	}
	\label{str}
\end{figure} 
%

%
\begin{figure*}[htp]
	\includegraphics[width=15cm]{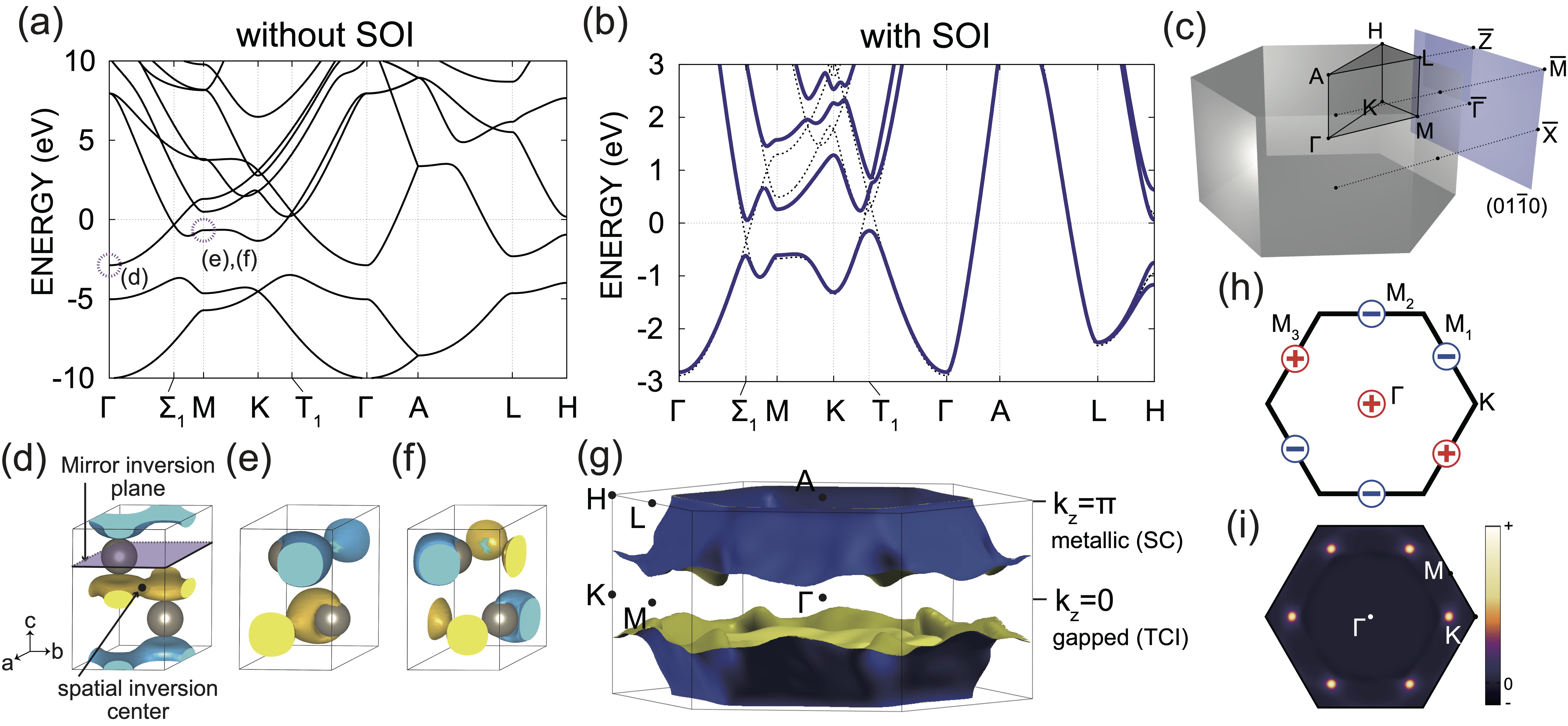}
	\caption{
		Electronic band structure.
		(a) Electronic band structure of Tl without the SOI. 
		(b) Electronic band structure of Tl with the SOI (purple solid line).
		That without the SOI is shown with black dotted lines.
		The energy is measured from the Fermi level.
		(c) Brillouin zone for bulk and the ($01\bar{1}0$) surface.
		(d)-(f) Eigenfunctions without the SOI at the time reversal invariant momenta (TRIM).
		The yellow and light blue represent the positive and negative components of the wave function, respectively.
		These eigenstates are marked in the band structure in (a).
		(d) corresponds to $\Gamma$ ($0,0,0$), (e) corresponds to $M_1$ ($1/2,0,0$) and $M_2$ ($0,1/2,0$), and (f) corresponds to $M_3$ ($-1/2,1/2,0$) shown in (h).
		(g) Fermi surface with the SOI.
		(h) Parity eigenvalue of each TRIM at $k_z=0$.
		(i) Berry curvature for mirror sector $+i$ on the $k_z=0$ plane.
	}
	\label{bulk}
\end{figure*} 
%


Therefore, it is interesting to consider whether there are such suitable
elements for the CMM
among the 25 elements which become superconducting under ambient pressure.
To make a topological gap in the band dispersion,
Tl and Pb are promising, because they have $6p$ valence electrons for which the spin-orbit interaction (SOI) is extremely strong.
We should also note that the gap opening must occur between two bands with different characters near the Fermi level.
While we found that the band structure of Pb is too simple to satisfy this condition, that of Tl is indeed topologically non-trivial.
In this Letter, we perform fully non-empirical calculation for the CMM considering superconductivity and surface relaxation, and show
that Tl has an ideal electronic state that harbors the CMM.

Tl is a nonmagnetic metal at room temperature, having a hexagonal close-packed (hcp) lattice with two lattice parameters $a$ and $c$ (Figs.~\ref{str}(a) and ~\ref{str}(b)). 
The space group of hcp is $P6_3/mmc$  (No. 194).
Unlike fcc elements, hcp elements have two equivalent atoms (A- and B-sites) in the unit cell.
At low temperature, the system becomes the type-I superconductor.
The transition temperature is relatively high among elemental superconductors, $T_c=2.39$ K~\cite{Eisenstein54}.
Tl changes to
type-II by doping 1-2 at\% of other $p$-elements.
The results for the hcp alloys are summarized in Appendix~\ref{sec:alloy}.

\section{Method}
	We calculate the band structures within density functional theory.
	The electronic structure is calculated using the generalized gradient approximation with and without the relativistic effect (GGA and GGA+SOI, respectively).
	We use  the \textit{ab initio} code OpenMX~\cite{openmx}.
	The valence orbital set is $s^3p^3d^3f^2$ for Tl 
	and the energy cutoff for the numerical integrations is 150 Ry for the calculation by OpenMX.
	We also use VASP~\cite{vasp} for the lattice optimization. 
	The energy cutoff is 50 Ry for the calculation by VASP.
	The lattice parameter $a$ and $c$ for Tl are 3.4566 and 5.5248 \AA ~\cite{Lide93}, respectively.
	The $12\times 12\times 8$ regular ${\bf k}$-mesh is employed for the bulk.
	
 We construct the $sp$ Wannier functions from the Kohn-Sham bands, using the maximally localized Wannier function~\cite{marzari97,souza01}.
	The density of states on the ($01\bar{1}0$) surface is calculated by the recursive Green's function method~\cite{turek97}.
	
 We optimize the lattice parameter of the $1 \times  24\times 1$ and $2 \times  6\times 2$ slabs of Tl (48 atoms)
	with $12\times 1\times 8$ and $6\times 1\times 4$ regular ${\bf k}$-mesh with vacuum in the ($01\bar{1}0$) direction with the GGA, respectively.
	We also optimize the lattice parameter of the $6 \times  6\times 2$ wire (108 atoms) with $1\times 1\times 8$ regular ${\bf k}$-mesh with vacuum perpendicular to the ($0001$) direction.
	We calculate the electronic band structure of the $1 \times  21\times 1$ slab (42 atoms) with the GGA+SOI.
 
	We calculate the phonon dispersion and electron-phonon coupling constant of hcp Tl and fcc Pb by using density functional perturbation theory code of quantum ESPRESSO~\cite{espresso}.
	We use the norm-conserved pseudopotential and the LDA-PZ exchange correlation functional~\cite{Perdew81}.
	The plane-wave cutoff for the Kohn-Sham orbital is set to 90 Ry.
	The superconducting gap calculation is performed by using SUPERCONDUCTING TOOLKIT in which the superconducting density functional theory (SCDFT) is implemented~\cite{Luders05,Marques05,Kawamura17}.
	Solving the SCDFT gap equation, we use $5\times5\times5$ {\bf q}-mesh for the phonon dispersion, $10\times10\times10$ {\bf k}-mesh for the Kohn-Sham orbital, and $40\times40\times40$ {\bf k}-mesh for the density of states, respectively~\cite{Kawamura17}.
 
	The density of states at the hinge is calculated by the recursive Green's function method from $1 \times  15\times 1$ (30 atoms) and $1 \times  9\times 1$ (18 atoms) slabs without and with the SOI, respectively.
	In the CMM calculation, we add the Zeeman field of 1.5 (4) meV
	for the surface Wannier function 
	at the transition point (in the TSC phase).

\section{Results}
We calculate the electronic structure of Tl by first-principles calculation (Figs.~\ref{bulk}(a) and ~\ref{bulk}(b))~\cite{openmx}.
Figure~\ref{bulk}(c) is the Brillouin zone.
First, we consider the case without the SOI (Fig.~\ref{bulk}(a)).
The electronic configuration of Tl is $6s^26p^1$.
The fully occupied bands from $-10$ to $-5$ eV originate from the $6s$ orbital.
The bands around the Fermi level originate from the $6p$ orbitals.
The top of the valence band originates from the $p_z$ orbital, which forms a covalent bond along the $z$ direction and stabilizes the structure as shown in Fig.~\ref{bulk}(d).
In the vicinity of the $M$ point, the $p_x$ and $p_y$ orbitals become the top of the valence band, forming a covalent bond in the $xy$ plane as shown in Figs.~\ref{bulk}(e) and ~\ref{bulk}(f).
In the $k_z = 0$ plane, the band structure is semimetallic.
The $p$ bands intersect and form a nodal line around the Fermi level.
The $p_z$ orbital and the $p_x$ and $p_y$ orbitals have
the odd ($-$) and even ($+$) mirror symmetry $M_z$ with respect to a plane perpendicular to the $c$-axis, respectively.
On the other hand, there are metallic bands near $k_z = \pi$.
All the bands at $k_z = \pi$ are doubly degenerate
owing to the crystal symmetry, where no bonding or anti-bonding state is formed between the A and B sites.

Figure~\ref{bulk}(b) shows the band structure with the SOI.
Tl is a $6p$ orbital system and has the strong SOI.
Figure~\ref{bulk}(g) shows the Fermi surface with the SOI.
The SOI breaks the degeneracy of the nodal line in the spinless system and open a large gap.
For example,
the size of the gap is about 0.8 eV at the $\Sigma _1$ point.

We investigate whether the band inversion at $k_z = 0$ is topological or not.
First we calculate the parity of the occupied bands (Fig.~\ref{bulk}(h)).
However, we find that the
$Z_2$ invariant
on the $k_z=0$ plane is 0, so that the $k_z=0$ plane does not correspond to a two-dimensional topological insulator.
Next we consider the mirror Chern number at $k_z = 0$.
The eigenfunction at $k_z=0$ is invariant to $M_z$ and has an eigenvalue $+i$ or $-i$ for $M_z$.
We calculate the topological invariant for each mirror sector from the Berry curvature,
\begin{align}
\bm{\Omega} ^{\pm i}({\bf k})=i\sum ^{\text{occ.}} _ {\pm i} \bm{\nabla } \times  \langle u^{\pm i}({\bf k}) | \bm{\nabla } | u^{\pm i}({\bf k}) \rangle ,
\end{align}
and get the mirror Chern number $|N_M|=|(N_{+i} - N_{-i})/2 |=2$.
Therefore, the gapped $k_z = 0$ plane in Tl corresponds to a topological crystalline insulator (TCI).

On the other hand, the Fermi surface around $k_z = \pi$ is simple and almost the same as that without the SOI.
This large Fermi surface also becomes gapped when the system undergoes a phase transition to $s$-wave superconductivity.
Therefore,
Tl is a promising candidate for the CMM
in which the topological crystalline gap and the $s$-wave superconducting gap coexist in the bulk (Fig.~\ref{bulk}(i)).

\begin{figure*}[ptb]
	\centering 
	\includegraphics[width=15cm]{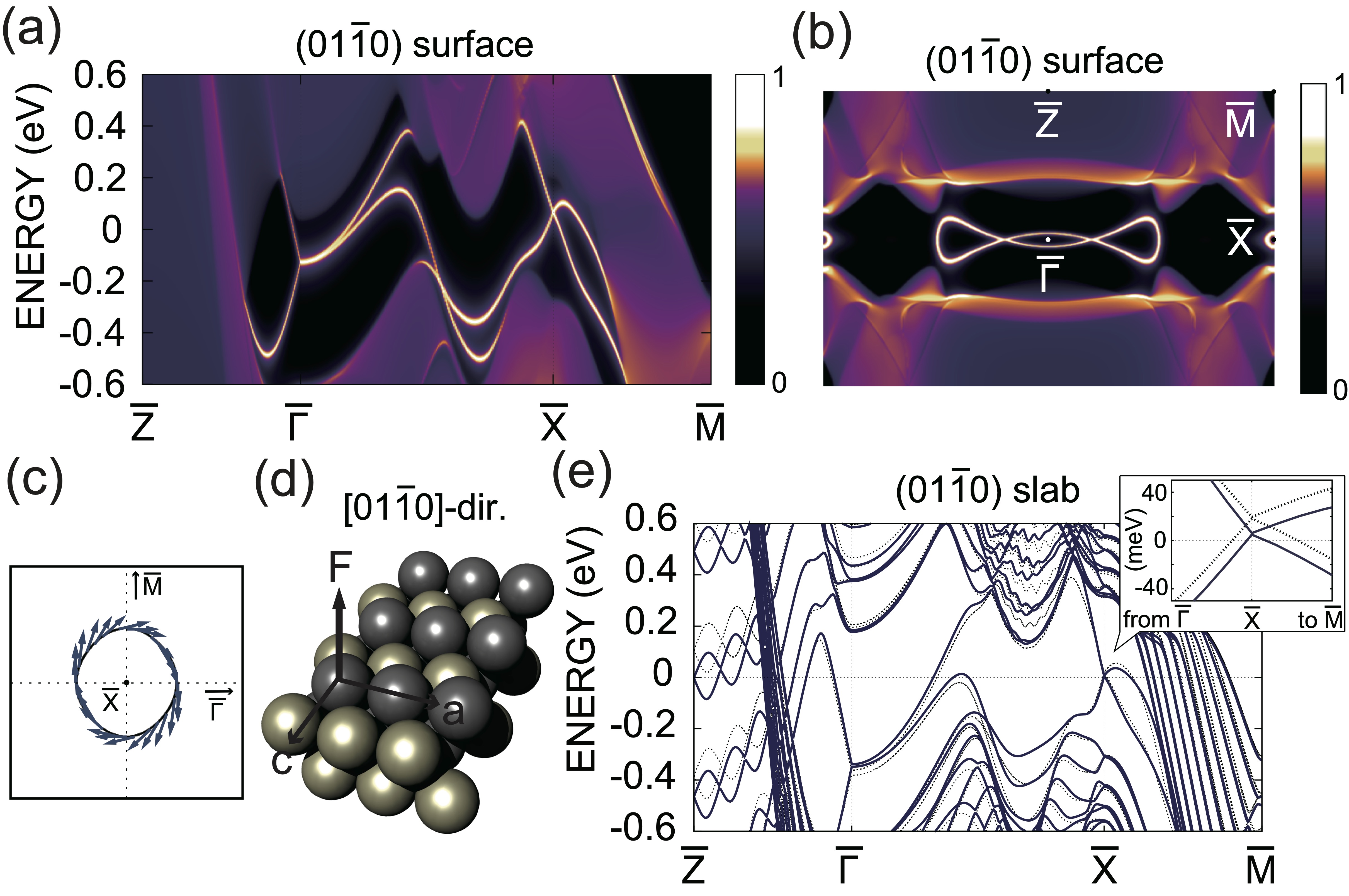} 
	\caption{Topological surface state.
		(a) Electronic band structure for the ($01\bar{1}0$) surface.
		(b) Charge distribution for the ($01\bar{1}0$) surface at the Fermi level.
		(c) Spin texture around the $\bar{X}$ point.
		(d) Electronic band structure of the  ($01\bar{1}0$) $1 \times  21\times 1$ slab (42 atoms).
		The blue solid line is the result with the lattice optimization.
		The black dotted line is the result without the lattice optimization for comparison.
		Inner window is the magnified figure around the $\bar{X}$ point near the Fermi level.
		The energy is measured from the Fermi level.
	}
	\label{surface}
\end{figure*}

Next we show the surface state of Tl for the ($01\bar{1}0$) direction.
The hcp structure forms a honeycomb lattice when viewed from the $c$ direction and the ($01\bar{1}0$) surface corresponds to the so-called zigzag edge of graphene (Fig.~\ref{str}(b)).
Unlike most elements having the hcp structure,
the most stable surface in Tl is not ($0001$) but ($01\bar{1}0$) ~\cite{Tuo15}.
This is due to the fact that the valence band mainly originates from the covalent bond of the $p_z$ orbital.
Figure~\ref{surface}(a) shows the electronic structure of the ($01\bar{1}0$) surface obtained from the recursive Green's function method~\cite{turek97} and Fig.~\ref{surface}(b) shows its charge distribution at the Fermi level.
The corresponding two-dimensional Brillouin zone is shown in Fig.~\ref{bulk}(c).
The ($01\bar{1}0$) surface holds the mirror symmetry $M_z$.
Reflecting the mirror Chern number $|N_M|$ is 2, two non-equivalent Dirac dispersions appear around the $\bar{\Gamma}$ and the $\bar{X}$ points on the surface.
The Dirac point has the Rashba-like spin texture as
in the surface state of the usual topological insulators (Fig.~\ref{surface}(c)).

We confirm the mirror symmetry at surface and the energy level of the Dirac point in the slab calculation.
In the process of structure optimization, the force acting on each atom is only in the direction perpendicular to the surface (Fig.~\ref{surface}(d)), and the mirror symmetry does not change~\cite{vasp}.
Figure~\ref{surface}(e) shows the electronic band structure of the slab Tl.
Due to the charge redistribution,
the position of the Dirac point
changes by 0.1 eV
in the slab calculation.
When considering the optimization, the energy level at the Dirac point at $\bar{X}$ gets lower slightly and is located almost at the Fermi level ($\sim 5$ meV).

\begin{figure*}[htp]
	\includegraphics[width=15cm]{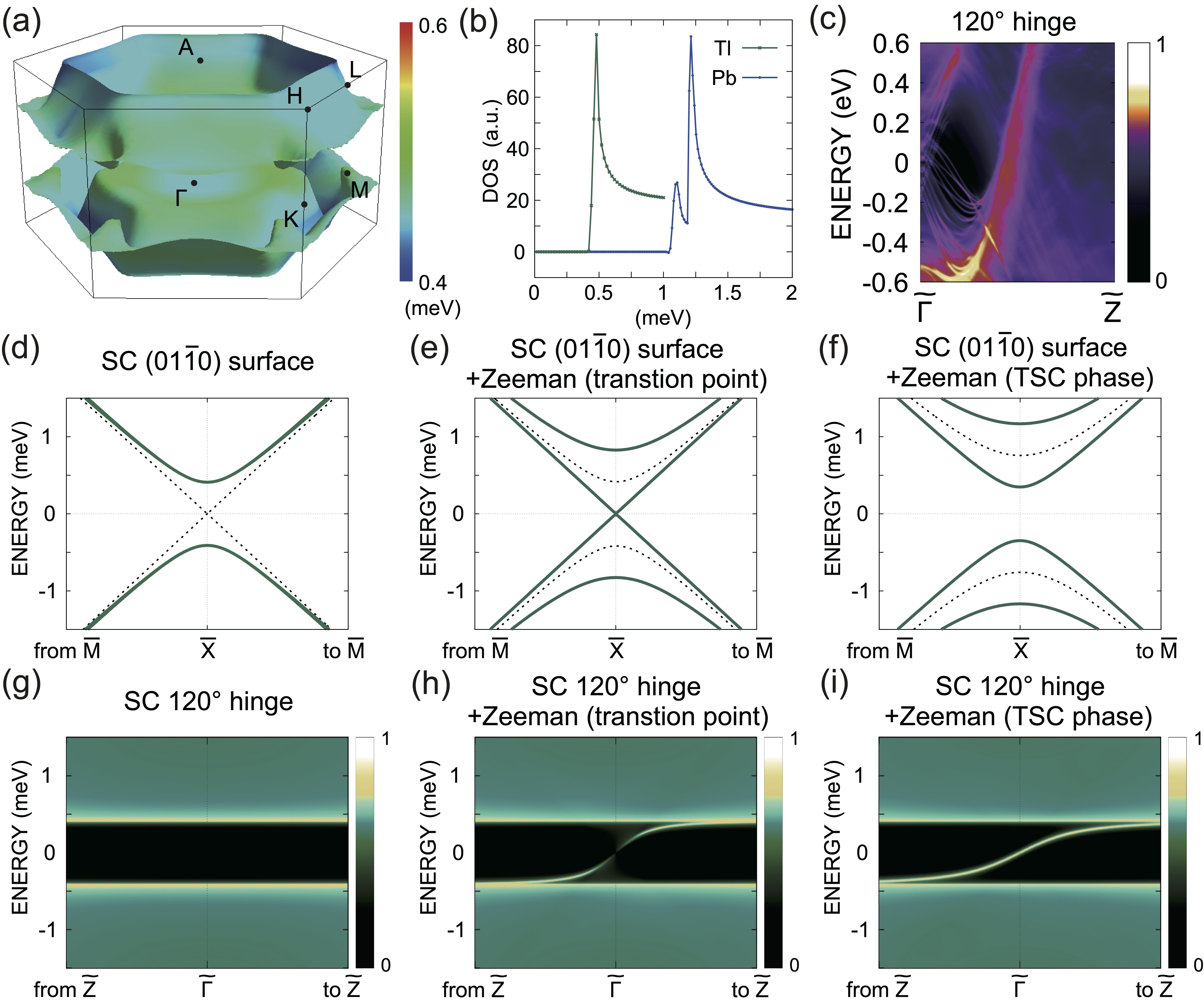}
	\caption{Superconducting (SC) state and chiral Majorana mode.
		(a) ${\bf k}$-dependence of the SC gap on the Fermi surface.  
		(b) Density of states in the SC phase.
		For comparison, we also show the result for fcc Pb. 
		(c) Electronic structure of the $120^\circ $ hinge of Tl in normal phase.
		(d) Electronic structure of the ($01\bar{1}0$) surface of Tl in the normal/SC phase (black dotted line/green solid line, respectively).
		(e)/(f) Electronic structure of the ($01\bar{1}0$) surface of Tl in the SC phase with the Zeeman field
		(green solid line) at the transition point from the trivial SC phase to the TSC phase/in the TSC phase, respectively.
		In addition to the SC phase, that in normal phase is shown in black dotted line.
		(g) Electronic structure of the $120^\circ $ hinge of Tl in the SC phase.
		(h)/(i) Electronic structure of the $120^\circ $ hinge of Tl in the SC phase with the Zeeman field,
		which corresponds to the phase of (e)/(f), respectively. 
		The energy is measured from the Fermi level.
	}
	\label{sc}
\end{figure*}

Here we discuss the superconducting phase.
We calculate the electron-phonon interaction from first-principles and determine the superconducting gap~\cite{espresso,Kawamura17,Luders05,Marques05}.
Figure~\ref{sc}(a) is the ${\bf k}$ dependence of the superconducting gap on the Fermi surface.
The ${\bf k}$ dependence of the superconducting gap is small.
The gap structure has no node, 
so that the pairing symmetry is $s$-wave.
Figure~\ref{sc}(b) is density of states in the superconducting phase.
We also calculate that of fcc Pb located next to Tl in the periodic table as a comparison.
While Pb has a double-peak gap structure which is observed in the recent experiment~\cite{Feldman17},
Tl has a simple gap structure.
The superconducting gap is 0.41 meV and the transition temperature is 2.7 K,
which is consistent with experimental value 2.39 K.

We show the emergence of the CMM at the hinge of Tl protected by the non-trivial topology of the surface state in the superconducting phase.
Figure~\ref{sc}(c) is the hinge state of Tl in the normal phase.
We calculate superconducting state at the surface (Figs.~\ref{sc}(d)-(f)) and the hinge (Figs.~\ref{sc}(g)-(i)).
By using the lattice optimization, we confirm that the hinge between the A-terminated $(01\bar{1}0)$ and the B-terminated $(10\bar{1}0)$ surfaces (Fig.~\ref{str}(b)) is stable.
In the superconducting state, we use the gap as a constant value since the ${\bf k}$ dependence and the frequency dependence near the Fermi level of the superconducting gap are small.
Here we calculate the Bogoliubov-de Gennes (BdG) Hamiltonian 
\begin{equation}
\begin{aligned}
H_{\text{BdG}}= &
\begin{pmatrix}
H({\bf k})-\mu       &    \Delta _{\bf k}        \\
    \Delta ^{\ast }_{\bf k}    &   -H^{\ast} (-{\bf k})+\mu
\end{pmatrix},
\label{BdG}
\end{aligned}
\end{equation}
where $H({\bf k})$ is the tight-binding Hamiltonian {of a slab Tl constructed from Wannier functions of the $6s$ and $6p$ orbitals,
$\mu$ is the chemical potential, and $\Delta _{\bf k}$ is the superconducting gap $\Delta _{\bf k} = i \Delta \sigma _{y}$.
Since the energy level shifts
due to the formation of the $(10\bar{1}0)$ surface and that the scale of 1 meV is smaller than the precision of the first-principles calculation with structure optimization,
we adjust the surface chemical potential so that the Dirac point is located at the Fermi level.
The results in which the Dirac point is slightly deviated from the Fermi level are essentially the same (see Appendix~\ref{sec:hinge}).
Figure~\ref{sc}(d) is the band structure of the slab in the superconducting phase.
In the same way as the bulk state, the topological surface state is gapped by the proximity effect.

We calculate the electronic structure of the hinge by using the recursive Green's function method on the slab (Fig.~\ref{sc}(g)).
The intensity above and below the superconducting gap includes the contributions of states other than the Dirac cone around the $\bar{X}$ point.
At this time, the hinge is completely gapped and the nontrivial chiral state does not appear.

Finally, we show the topological phase transition at the surface by the Zeeman field.
Experimentally, this can be achieved by attaching ferromagnetic insulators.
We introduced the Zeeman field perpendicular to the surface for just
one layer from the $(01\bar{1}0)$ surface,
which keeps the energy difference between the ${\bf k}$ and $-{\bf k}$ states in the normal phase to zero.
When the Zeeman field is increased, the gap on the surface closes once at the $\bar{\Gamma}$ point (Fig.~\ref{sc}(e)),
and the system becomes
topological superconducting (TSC) phase (Fig.~\ref{sc}(f)).

In Fig.~\ref{sc}(h) and (i), we show
the corresponding CMM formed by the Zeeman field for the surface.
Majorana fermions are particles which are also their own anti-particles.
In the superconducting phase, electrons and holes have symmetry of $\gamma (\epsilon )  = \gamma ^{\dagger}  (-\epsilon ) $.
At the Fermi level, $\gamma (0)  = \gamma ^{\dagger}  (0) $ is satisfied, and the particle and anti-particle coincide.
At the transition point, a pair of the CMMs emerges at the Fermi level (Fig.~\ref{sc}(h)).
This CMM has spread throughout the surface, and the surface intensity of the CMM at the $\tilde{\Gamma}$ point is 0 as shown in Fig.~\ref{sc}(h).
In the non-trivial superconducting phase, the CMM is localized at the hinge (Fig.~\ref{sc}(i)). 
Since the CMM is a nonlocal state represented by a half of the component of the normal wave function,
the CMM does not disappear unless it is mixed with another CMM passing through the $\tilde{\Gamma}$ point.

\section{Conclusion}
To summarize, we find that hcp Tl is a promising material for realizing the elusive chiral Majorana fermion.
The $k_z=0$ plane corresponds to the TCI with mirror Chern number $|N_M|=2$.
The surface and edge structures are stable.
Two inequivalent Dirac dispersions appear on the ($01\bar{1}0$) surface reflecting $|N_M|=2$ and one of the Dirac points is located almost at the Fermi level.
Tl is a textbook-like $s$-wave superconductor and the gap function has no significant wave-number dependence. 
Only one of the two Dirac points is relevant for the gap opening due to the superconducting transition,
and the CMM appears at the hinge under the Zeeman field
which can be exploited in the future quantum technology.
Since Tl has topological surface states protected by the mirror symmetry,
Tl would also show the MBS in magnetic fluxes characterized by $Z$~\cite{PhysRevLett.112.106401,PhysRevB.90.235141,PhysRevLett.125.136802}.
Therefore, Tl is a good platform for the topological superconductors and the Majorana states.

\begin{acknowledgments}
We thank Masatoshi Sato for a fruitful discussion.
M.H. acknowledges financial support from JST PRESTO (JPMJPR21Q6) and JST CREST (JPMJCR19T2 and JPMJCR23O3).
\end{acknowledgments}

\appendix

\section{Tl Alloy}
\label{sec:alloy}

In Fig.~\ref{alloy}, we show the results for Tl alloys.
We optimize the lattice parameter for In, Hg, and Pb in hcp structure using VASP.
The optimized lattice parameter $a$ ($c$) is 3.4646 (5.3456) \AA \ for In, 3.6092 (5.5939) \AA \ for Hg, and 3.5266 (5.8844) \AA \ for Pb, respectively.
We calculate the band structure of hcp Tl-In alloy by linear interpolation using the Wannier function and that of Hg-Tl and Tl-Pb alloys by virtual crystal approximation.

\begin{figure*}[htp]
	\includegraphics[width=12cm]{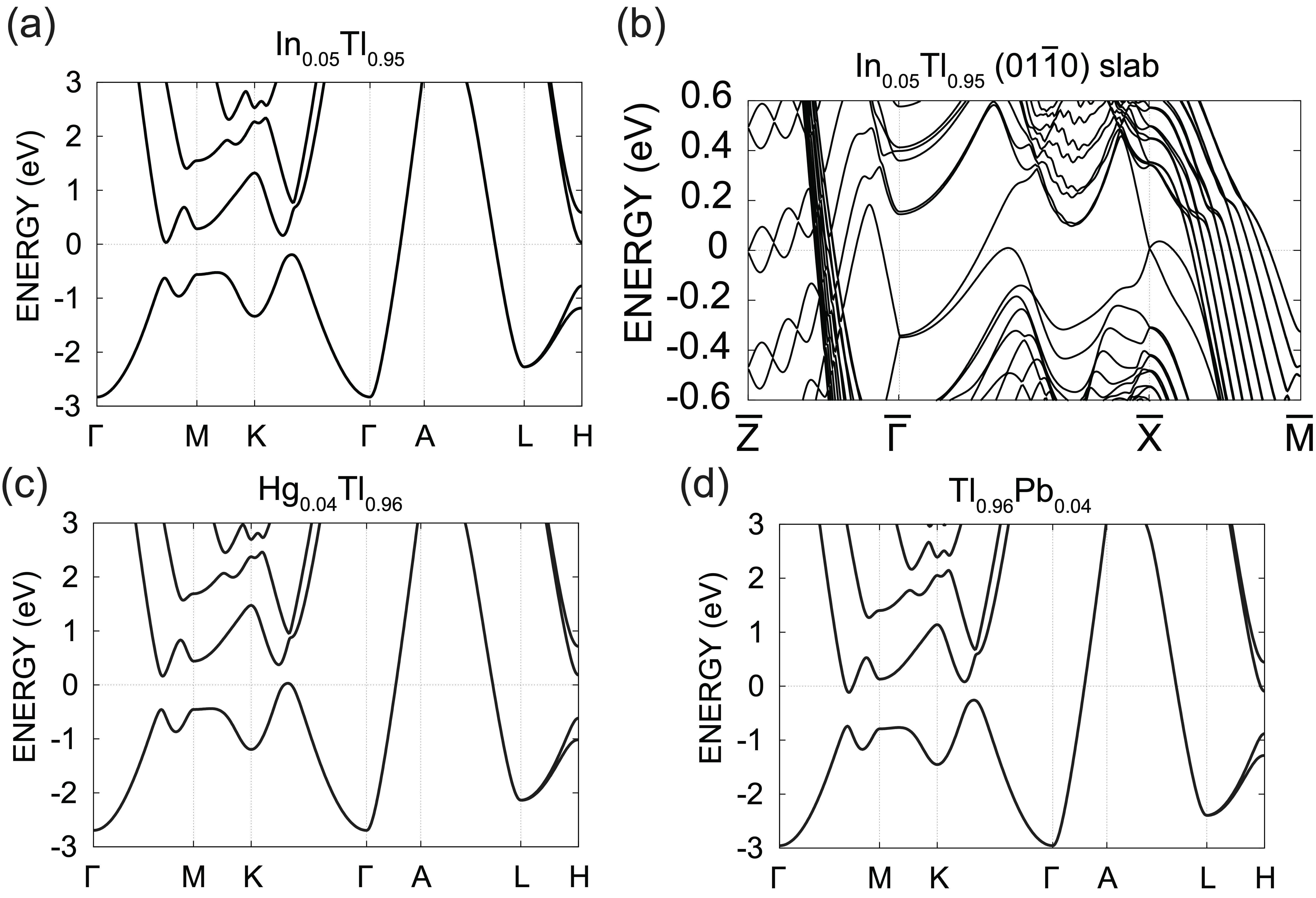}
	\caption{In-Tl alloy.
		(a) Electronic band structure of In$_{0.05}$Tl$_{0.95}$ alloy.
		(b)  Electronic band structure for the ($01\bar{1}0$) surface of In$_{0.05}$Tl$_{0.95}$ alloy.
		The size of the ($01\bar{1}0$) slab is $1 \times  21\times 1$ unit cell (42 atoms).
		(c) Electronic band structure of Hg$_{0.04}$Tl$_{0.96}$ alloy.
		(d) Electronic band structure of Tl$_{0.96}$Pb$_{0.04}$ alloy.
		The energy is measured from the Fermi level.
	}
	\label{alloy}
\end{figure*}

\section{Hinge state with the Zeeman field.}
\label{sec:hinge}

Figures~\ref{sc5meV}(a)-(c) are the surface band structures in the superconducting phase when the chemical potential of the Dirac point deviates from the Fermi level.
The band is completely gapped.
Reflecting the position of the Dirac point in the normal phase, two Dirac points appear at the $\bar{X}$ point in the superconducting phase due to the particle-hole symmetry.
Although the smallest gap exists around the $\bar{X}$ point in the nonmagnetic phase, when the Zeeman field is increased, only the energy near the $\bar{X}$ point having the spin degeneracy changes significantly.
The band touching between the valence and conduction bands occurs at the $\bar{X}$ point.
When the Zeeman field is further increased, a gap opens and 
the surface superconducting state becomes topological.
The figures~\ref{sc5meV}(d)-(f) show the corresponding hinge band structure.
The shape of the band structure is qualitatively the same as that where the Dirac point is at the Fermi level.

\begin{figure*}[htp]
	\includegraphics[width=12cm]{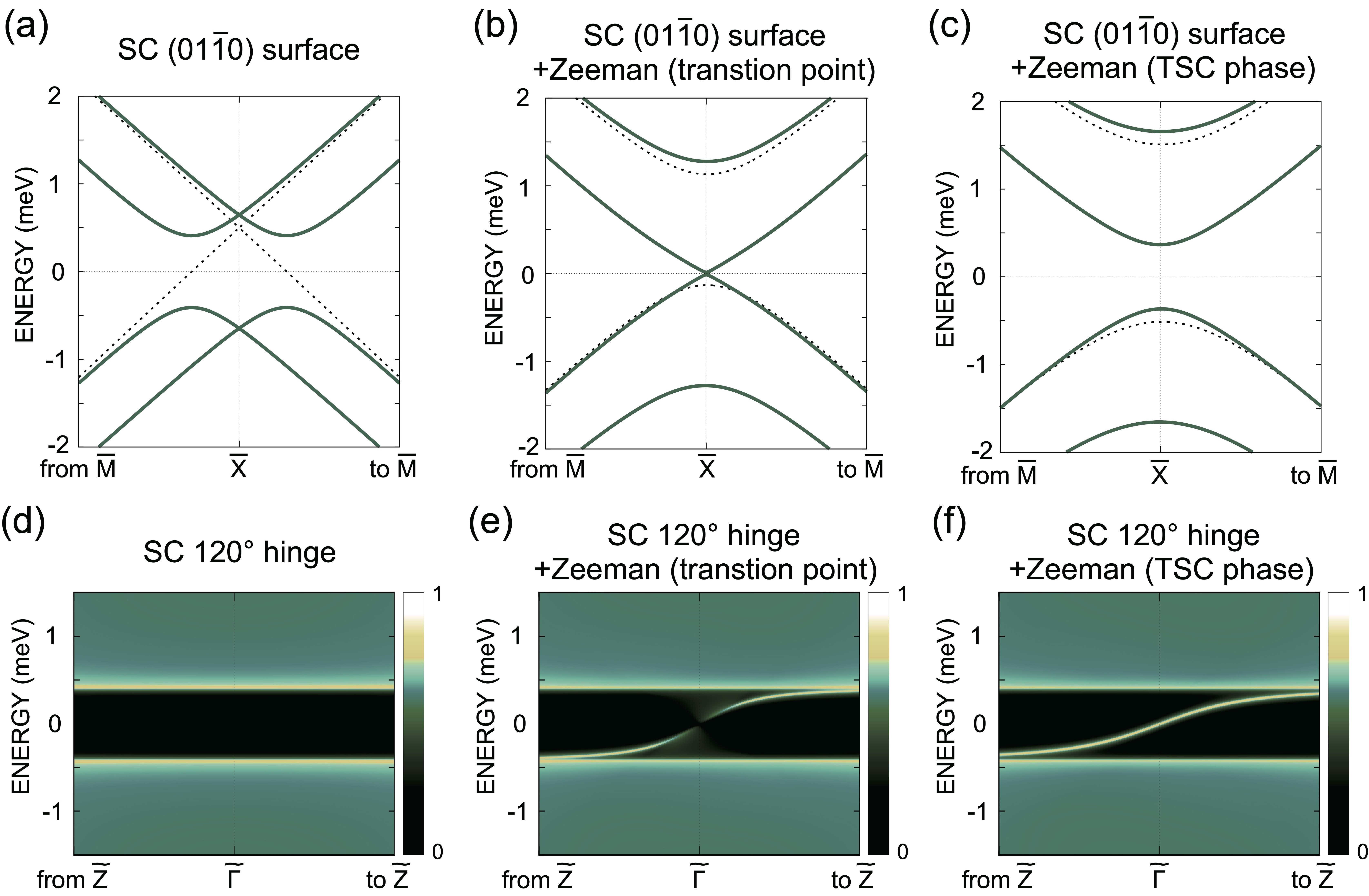}
	\caption{Superconducting (SC) state and chiral Majorana mode with the Dirac point at $\epsilon = 0.5$ meV.
		(a) Electronic structure of the ($01\bar{1}0$) surface of Tl in the normal/SC phase (black dotted line/green solid line, respectively), where the Dirac point is at $\epsilon = 0.5$ meV.
		(b)/(c) Electronic structure of the ($01\bar{1}0$) surface of Tl in the SC phase with the Zeeman field
		(green solid line) at the transition point from the trivial SC phase to the TSC phase/in the TSC phase, respectively.
		In addition to the SC phase, that in the normal phase is shown in black dotted lines.
		(d) Electronic structure of the $120^\circ $ hinge of Tl in the SC phase.
		(e)/(f) Electronic structure of the $120^\circ $ hinge of Tl in the SC phase with the Zeeman field,
		which corresponds to the phase of (b)/(c), respectively. 
		The energy is measured from the Fermi level.
	}
	\label{sc5meV}
\end{figure*}

\bibliography{full}

\end{document}